\documentclass[a4paper,fleqn]{cas-sc}
\usepackage[numbers]{natbib}
\usepackage[misc]{ifsym} 

\usepackage{subfigure}
\usepackage{graphicx}
\usepackage{caption}
\usepackage{booktabs} 

\def\tsc#1{\csdef{#1}{\textsc{\lowercase{#1}}\xspace}}
\tsc{WGM}
\tsc{QE}
\tsc{EP}
\tsc{PMS}
\tsc{BEC}
\tsc{DE}

\begin{document}
\begin{sloppypar}
	\def\textpagefraction{.001}
	\shorttitle{}
	
	\title [mode = title]{Self-supervised Noise2noise Method Utilizing Corrupted Images with a Modular Network for LDCT Denoising}

	
	\author[mymainaddress]{Yuting Zhu}
	\author[mymainaddress]{Qiang He}
	

	\author[mythirdaryaddress]{Yudong Yao}
	
	\author[mymainaddress, mysecondaryaddress]{Yueyang Teng\corref{mycorrespondingauthor}}
	\cortext[mycorrespondingauthor]{Corresponding author}
	\ead{tengyy@bmie.neu.edu.cn}
	
	\address[mymainaddress]{College of Medicine and Biological Information Engineering, Northeastern University, Shenyang  110169, China}
	\address[mysecondaryaddress]{Key Laboratory of Intelligent Computing in Medical Image, Ministry of Education, Shenyang  110169, China}
	\address[mythirdaryaddress]{Department of Electrical and Computer Engineering, Stevens Institute of Technology, Hoboken, NJ 07102, USA}

\begin{abstract}
Deep learning is a very promising technique for low-dose computed tomography (LDCT) image denoising. However, traditional deep learning methods require paired noisy and clean datasets, which are often difficult to obtain. This paper proposes a new method for performing LDCT image denoising with only LDCT data, which means that normal-dose CT (NDCT) is not needed. We adopt a combination including the self-supervised noise2noise model and the noisy-as-clean strategy. First, we add a second yet similar type of noise to LDCT images multiple times. Note that we use LDCT images based on the noisy-as-clean strategy for corruption instead of NDCT images. Then, the noise2noise model is executed with only the secondary corrupted images for training. We select a modular U-Net structure from several candidates with shared parameters to perform the task, which increases the receptive field without increasing the parameter size. The experimental results obtained on the Mayo LDCT dataset show the effectiveness of the proposed method compared with that of state-of-the-art deep learning methods. The developed code is available at \href{https://github.com/XYuan01/Self-supervised-Noise2Noise-for-LDCT}{https://github.com/XYuan01/Self-supervised-Noise2Noise-for-LDCT}.

\end{abstract}
	
	
	\maketitle
\section{Introduction}
Computed tomography (CT) has been widely used in preventive medicine and disease screening to obtain internal information about patients. As CT technology has become more widespread, concerns about its radiation risks have increased, and the ``as low as reasonably achievable'' (ALARA) principle has been accepted by physicians and patients to minimize radiation doses \cite{smith2009radiation}. Clinically, low-dose CT (LDCT) is used for examinations to reduce the risk of radiation exposure to patients; however, this leads to more noise, which can adversely affect diagnoses to some extent. Reducing the noise in LDCT images has been a hot topic in medical imaging in recent years.

Many methods have been proposed to solve this problem. These methods are usually divided into three main categories: 1) iterative tomographic reconstruction; 2) sinogram domain filtering; and 3) image postprocessing methods. Most traditional iterative tomographic reconstruction methods are dedicated to improving the reconstructed results based on prior information; an example is the iterative asymptotic minimum variance method (SAMV) \cite{abeida2012iterative}. However, they usually have massive computational costs. The essence of sinogram domain filtering methods, for example, the inverse fast Fourier transform (IFFT) \cite{schomberg1995gridding} and the bayesian Poisson denoising algorithm based on nonlocal means and stochastic distances \cite{evangelista2022new}, is to reconstruct a clear image by processing the original projection data. In some situations, sinogram domain filtering may suffer from the risk of introducing global noise in the image domain. Compared to iterative reconstruction and sinogram domain filtering, image postprocessing approaches do not require access to raw data and additional hardware costs, making them easier to incorporate into the imaging process and causing them to attract increasing attention; examples include the block matching 3D (BM3D) method \cite{feruglio2010block} and the bilateral filtering method \cite{manduca2009projection}.

In recent years, deep learning methods have emerged and gained widespread attention; these techniques are image postprocessing methods. Significant achievements have been made in denoising tasks using these methods, which can be classified into supervised, unsupervised, and self-supervised approaches depending on the type of input data. Supervised methods perform the task by inputting noise-free data into the learning noise distribution, and they can usually achieve good results. For example, the denoising convolutional neural network (DnCNN) \cite{zhang2017beyond} performs denoising by learning the residuals between noise-free data, borrowing the residual learning method, and combining it with batch normalization to improve the training speed and denoising performance of the constructed model. U-Net \cite{ronneberger2015u} fuses the low-level feature map with the corresponding high-level feature map, and its completely symmetric U-shaped structure makes the process of fusing the front and back features more thorough, yielding an increase in the low- and high-resolution information in the target image. It has been used for LDCT denoising tasks by combining multi-feature channel attention module \cite{zhang2023novel}. UNet++ \cite{zhou2018unet++} enhances the effect of U-Net by capturing features at different levels and integrating them via the feature overlay technique. It adds U-Net structures with different sizes to enhance the feature fusion process. A convolutional neural network combining wavelet transform and subpixel convolution have also been proposed for LDCT denoising tasks \cite{li2022adaptive}. Unsupervised deep learning methods that do not require paired images, for example, generative adversarial networks (GANs) \cite{goodfellow2020generative}, cycle-consistent GANs (CycleGANs) \cite{zhu2017unpaired}, Dual-GANs \cite{yi2017dualgan}, and artifact disentanglement networks (ADNs) \cite{liao2019adn}, have also been widely studied. A GAN model combines a discriminative network and a generative network so that the two form an adversarial process to improve their effectiveness. A CycleGAN has two discriminators and two generators, which combine a cycle consistency loss and an adversarial loss to form the whole model. Such models have been used for LDCT denoising tasks \cite{wolterink2017generative}. Some new CycleGANs with one generator \cite{gu2021adain} or combining the the similarity-guided discrimination loss \cite{zhao2023dual} has also been proposed for LDCT denoising tasks. A Dual-GAN performs an image domain transformation by combining dual learning and GAN networks with two discriminators and two generators. In fact, these GAN-type networks use a set to guide the imaging procedure to another set instead of pairwise data. ADNs are also unsupervised methods that have achieved good results. An ADN separates metallic noise in the input image by introducing an inductive bias and combining an adversarial loss, a reconstruction loss, an artifact consistency loss, and a self-reduction loss. Thus, both supervised and unsupervised learning methods require noisy and clean data, which are still limited in certain situations.

To solve the above problem, self-supervised deep learning methods have been proposed. These approaches do not require targeted images at all; they usually form data pairs by adding additional noise or creating blind spots to corrupt the original data, making it much less difficult to obtain data. The noisy-as-clean \cite{xu2020noisy} technique aims to improve the accuracy of denoising models by training them to reconstruct clean images from only noisy inputs. This strategy can greatly improve the utilization of noisy images. Furthermore, many advanced self-supervised denoising methods have been proposed. Noise2noise \cite{lehtinen2018noise2noise} (N2N) applies different noise patterns to the same clean image; then, a large number of noisy image pairs can be generated for training. Noise2void \cite{krull2019noise2void} (N2V) assumes that each pixel point in the given image does not exist independently and performs denoising by adding a blind spot to the central pixel of each patch, preventing the model from becoming a direct mapping of that pixel point. Blind2unblind \cite{wang2022blind2unblind} (B2U) makes blind spots visible by adding them multiple times at different locations and predicts pixels by learning the context around the blind spots to achieve denoising. Neighbor2neighbor \cite{huang2021neighbor2neighbor} (NBR2NBR) obtains data pairs by subsampling noisy images and uses these noisy data pairs to perform denoising based on N2N.

This paper proposes a self-supervised N2N model based on the abovementioned noisy-as-clean strategy for LDCT image denoising. First, we build a dataset, which repeatedly adds noise with an independent distribution and a mean of zero to the given LDCT images. It is worth noting that the process of adding noise to the LDCT images is based on the noisy-as-clean strategy, and repeatedly adding noise forms a dataset that satisfies the self-supervision requirement of N2N. Then, a modular network structure is adopted, in which several modules share the same structure and parameters and then form the network by linking in turn. Such a network can not only improve the receptive field but also reduce the parameter scale. Through experiments, we select U-Net from several candidates as the backbone module. The experimental results also show the effectiveness and feasibility of the proposed method in comparison with the state-of-the-art methods.

\section{Methodology}

\begin{figure*}
        \centering
	\includegraphics[width=1\textwidth]{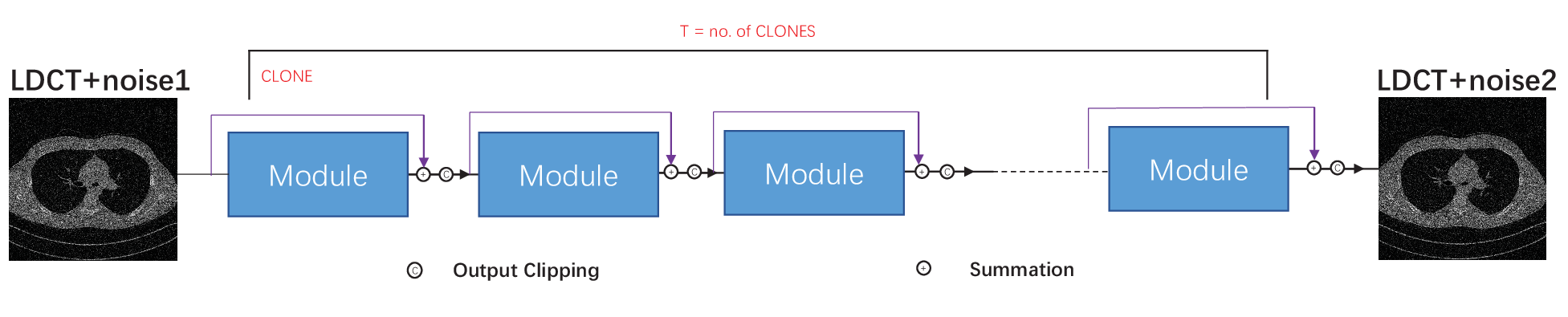}
	\caption{Overall network architecture.}
	\label{structure}
	\vspace*{2ex}
\end{figure*}

This paper constitutes the overall structure of the proposed method by combining the self-supervised N2N model and the noisy-as-clean strategy, which is performed with a modular network structure. The overall network structure is shown in Figure \ref{structure}.
 \subsection{Self-supervised noise2noise}
The task of denoising LDCT images with supervised learning can be simply formulated as:
$$\underset{\theta}{argmin}\sum_{i}^{}{L(f_{\theta}(\hat{x}_{i}),y_{i})}\mathrm{},\eqno(1)$$
 where $\hat{x_{i}}$ represents
corrupted inputs, $y_{i}$ represents clean targets, $f_{\theta}$ stands for the denoising function with optimal parameters $\theta$ and $L$ denotes the loss function.
The supervised learning method is trained to solve the point estimation problem by minimizing its loss function. The loss function can be $L_{2}$, where the minimum is found at the arithmetic mean of the observations, or $L_{1}$, where the minimum is found at the median of the observations.

The training task for a set of input-target pairs $(\hat{x_{i}},y_{i})$, where $x = y + n_{o}$ ($n_{o}$ is the observed noise), is a minimization problem with the following form:
 
 $$\underset{\theta}{argmin}\mathbb{E}_{x}\left \{ \mathbb{E}_{y|x}\left \{ {L(f_{\theta}({x}),y)} \right \}  \right \} .\eqno(2)$$
 
 Supervised learning can solve this problem separately for each clean image by minimizing the loss function.
 
 The usual process of training regressors by Equation (1) over a finite number of input-target pairs $({x_{i}},y_{i})$ implies a 1:1 mapping between the inputs and targets. However, in reality, the mapping is multiple-valued due to the highly complex distribution $p(y|x)$. When a neural network regressor is trained using low-/high-noise image pairs and the $L_{2}$ loss, the network learns to output the average of all plausible explanations. However, for certain problems, this tendency can be beneficial. Utilizing the property that $L_{2}$ minimizations are unchanged if we replace the targets with random numbers whose expectations match those of the targets, Lehtinen \emph{et~al.} \cite{lehtinen2018noise2noise} corrupted the training targets of a neural network with zero-mean noise without changing what the network learned. This leads to the following empirical risk minimization task: 
 
 $$\underset{\theta}{argmin}\sum_{i}^{} L(f_{\theta}(\hat{x}_{i}),\hat{y}_{i}),\eqno(3)$$
 where both the inputs and the targets are drawn from a corrupted distribution. An explicit likelihood or density model is not needed as long as the data are distributed according to these models.

 \subsection{Noisy-as-clean strategy}
The natural noise generated during LDCT imaging is dependent on the signal, and to avoid the problem that the image prior and noise statistics are different in the training and testing sets, Xu \emph{et~al.} \cite{xu2020noisy} used the noisy-as-clean strategy to train a self-supervised network. In this strategy, the corrupted image serves as the "clean" target, and then the corrupted and secondary corrupted images can be used as a pair for training.

The authors studied a practical assumption on noise statistics: the expectation $\mathbb{E}[x]$ and variance $Var[x]$ of the signal intensity are much stronger than those of the noise $\mathbb{E}[n_{o}]$ and $Var[n_{o}]$ (which are negligible but not necessarily zero). This is effective for LDCT denoising because the tissue in normal-dose CT (NDCT) is observed with almost no noise effect. The noisy image $x$ should have a similar expectation to that of the clean image $y$.

A noise model $n_s$ is added to the observed noisy image $x$, forming a new noisy image $z=x+n_{s}$. It is assumed that $n_{s}$ is statistically close to $n_{o}$; i.e.,
$\mathbb{E}[n_{s}]\approx
\mathbb{E}[n_{o}]$, and $Var[n_{s}]\approx Var[n_{o}]$. The simulated noisy image $z$ has a similar expectation to that of the observed noisy image $x$.
By the law of total expectation \cite{billingsley2008probability}, we have

$$\mathbb{E}_{x}[\mathbb{E}_{z}[z|x]]=\mathbb{E}_{z}\approx \mathbb{E}_{x}=\mathbb{E}_{y}[\mathbb{E}_{x}[x|y]].\eqno(4)$$

Since the $L_{2}$ loss function and the conditional probability density functions $p(x|y)$ and $p(z|x)$ are all continuous everywhere, the optimal network parameters $\theta ^{*} $ of Equation (3) change little with the addition of negligible noise $n_{o}$ or $n_{s}$. When $\mathbb{E}_{x|y}\left\{{L(f_{\theta}({x}),y)}\right\}$ according to the $y$-conditioned expectation is replaced with the $x$-conditioned expectation of $\mathbb{E}_{z|x}\left\{{L(f_{\theta}({z}),x)}\right\}$, $f_{\theta}$ obtains similar $y$-conditioned optimal parameters $\theta ^{*} $ to those of Equation (2), which are equivalent to
:

$$\underset{\theta}{argmin}\mathbb{E}_{z}\left \{ \mathbb{E}_{z|x}\left \{ {L(f_{\theta}({z}),x)} \right \}  \right \}. \eqno(5)$$

In combination with the self-supervised framework, the task of empirical risk minimization becomes

 $$\underset{\theta}{argmin}\sum_{i}^{} L(f_{\theta}(\hat{z}_{i}),\hat{x}_{i}),\eqno(6)$$
 where $\hat{z}_{i}$ and $\hat{x}_{i}$ are both corrupted LDCT images.

However, when we reduce the radiation dose according to the ALARA principle, electronic noise $n_{o}$ is inevitably generated during CT imaging, which usually follows a combined Poisson and Gaussian distribution. Furthermore, the framework requires zero-mean noise, so combined zero-mean Poisson and Gaussian noise is added.

As our aim is LDCT denoising, the noise can be modeled by mixing the Poisson and Gaussian noise distributions. Fortunately, both distributions are linearly additive; i.e., the additional variable of two Poisson (or Gaussian) distributed variables is still Poisson (or Gaussian) distributed. Assume that the observed noise $n_{o}$ and simulated noise $n_{s}$ are $x$-dependent and $z$ dependent, respectively. Let the Poisson distribution be parameterized by $\lambda_{o}$ and $\lambda_{s}$, and denote the Gaussian distributions as $\mathcal{N}(0,\sigma _{o}^{2}) $ and $\mathcal{N}(0,\sigma _{s}^{2}) $, i.e.,

$$n_{o}\sim x\odot \mathcal{P} (\lambda _{o})+\mathcal{N}(0,\sigma _{o}^{2}),$$$$n_{s}\sim z\odot \mathcal{P} (\lambda _{s})+\mathcal{N}(0,\sigma _{s}^{2})$$$$\hspace{1.5em} \approx x\odot \mathcal{P} (\lambda _{s})+\mathcal{N}(0,\sigma _{s}^{2}),\eqno(7)$$
where $x\odot \mathcal{P} (\lambda _{o})$ and $z\odot \mathcal{P} (\lambda _{s})$ indicate that the noises $n_{o}$ and $n_{s}$ are elementwise dependent on $x$ and $z$, respectively. To this end, we have that $$n_{o}+n_{s}\sim x\odot \mathcal{P} (\lambda _{o}+\lambda _{s})+\mathcal{N}(0,\sigma _{o}^{2}+\sigma _{s}^{2}+2\rho \sigma _{o}\sigma _{s}),\eqno(8)$$
where $\rho $ is the correlation between $n_o$ and $n_s$. Here, $\rho $ = 0 because the noises are independent. This indicates that the summed noise variable $n_{o} + n_{s}$ still follows a mixed $x$-dependent Poisson and Gaussian distribution, guaranteeing the consistency of the noise statistics between the observed realistic noise and the simulated noise.

 \subsection{Modular network structure}
The quality of CT imaging gradually improves as the radiation dose gradually increases from low to high. To a significant degree, the reconstruction process from low-quality LDCT images to high-quality NDCT images can be gradually simulated by neural networks. We design a modular network for LDCT image denoising in which all modules share the same structure and parameters. Each module improves the image quality by a small increment and constitutes a stepwise improvement of the denoising task, and all these modules together produce a series of denoised images. At the same time, the addition of modules increases the model capacity of the whole network without introducing new parameters. Formally, our network can be formulated as follows:

$$I_{den}=f^T(I_{LDN})=\underbrace{(f\circ f\circ \cdots \circ)f}_{T=\#f}(I_{LDN}),\eqno(9)$$
where $I_{den}$ and $I_{LDN}$ denote a denoised image and a noisy image, respectively. The operator $f$ denotes a functional composition operation, and $f^T$ denotes the $T$-fold product of the module. The parameters of all the modules are shared.

We select several advanced networks as candidate modules based on the subsequent experiments, including U-Net \cite{ronneberger2015u}, a recurrent residual convolutional neural network based on U-Net \cite{alom2018recurrent} (R2U-Net), an attention U-Net \cite{oktay2018attention} (AttU-Net), an attention-based recurrent residual U-Net \cite{zuo2021r2au} (R2AU-Net), a conveying-path-based convolutional encoder-decoder network \cite{shan20183} (CPCE), and a residual network \cite{he2016deep} (ResNet). U-Net uses a U-shaped network structure for upsampling and downsampling while retaining structural details using skip connections. R2U-Net adds a residual module and a recurrent module to U-Net, and the network deepens while avoiding gradient unlearning. AttU-Net adds a self-attention module to U-Net, making the skip connections more selective. R2AU-Net adds a residual module, a cyclic module, and a self-attention module to U-Net. CPCE uses convolution and deconvolution to construct its encoder-decoder structure and connects the ReLU activation function after each layer. ResNet achieves improved model performance by adding residual links between each pair of layers. It has been used for LDCT noise reduction tasks by combining the fractional-order total variation loss \cite{chen2021low}.


\section{Experiments}
We first introduce the configurations of the dataset and experiments. Then, the impacts of different backbone networks, different noise levels, and various numbers of modules on the resulting model performance are compared. We also discuss the effectiveness and necessity of the noise distribution and the noisy-as-clean strategy through ablation experiments. Finally, we compare our approach with several state-of-the-art self-supervised and unsupervised methods.
\subsection{Experimental description}
The dataset provided by the Mayo Clinic, called the NIH-AAPM-Mayo Clinic Low-Dose CT Grand Challenge, contains NDCT images and their corresponding simulated paired LDCT images, and it was used to evaluate the proposed method. The training set contains 3595 LDCT images of different tissues from 11 patients. The testing set was randomly selected from 8 unused patients, including 1198 CT images. It is worth mentioning that our method did not require NDCT images throughout the process, and the LDCT images were normalized before being input into the network.

The model was implemented in PyTorch on a PC (GPU NVIDIA Quadro P2200). The training process was optimized using the ADAM optimizer. In ADAM, $\beta _{1}$ was set to 0.9, $\beta _{2}$ was set to 0.99 and $\varepsilon $ was set to $10^{-8}$. The initial learning rate was $10^{-4}$ and was halved after every 20 epochs. Sixty epochs were used in total. The batch size was set to 4. Poisson noise and Gaussian noise (with a mean of 0 and a variance of 15) were added as a mixed noise model, which is shown in Equation (8). The intensity of the Poisson noise is only related to the pixel value. Its intensity increases as the light becomes stronger, meaning that the larger the pixel value of the image is, the more frequently Poisson noise appears. Here, we satisfied the condition of the self-supervised N2N framework by applying only zero-mean processing to the Poisson noise. Therefore, we next discuss only the variance of the Gaussian noise.

We used the peak signal-to-noise ratio (PSNR) \cite{huynh2008scope} and structural similarity index measure (SSIM) \cite{wang2004image} to objectively evaluate the proposed method.

The PSNR measures the denoising performance of a method by calculating the overall difference between the denoised LDCT image and the original NDCT image. A higher value represents better image quality. $$PSNR=10\log_{10}{(\frac{max^{2}}{\frac{1}{n} {\textstyle \sum_{i=1}^{n}(x_{i}-y_{i})^{2}}  } )} \eqno(10)$$, where $x_{i}$ and $y_{i}$ are the pixels of the LDCT and NDCT images, respectively, $n$ is the number of pixels in the image, and $max$ represents the maximum value of the image pixels.

The SSIM evaluates image quality by considering the overall structure of the given image. The value lies in the range of [0, 1], and the higher the value is, the closer the image structure is to that of the NDCT image.
$$SSIM(x,y)=\frac{(2\mu _{x}\mu _{y}-c_{1})(\sigma _{xy}+c_{2})}{(\mu _{x}^{2}+\mu _{y}^{2}+c_{1})(\sigma _{x}^{2}+\sigma _{y}^{2}+c_{2})} \eqno(11)$$, where $\mu _{x}$ and $\mu _{y}$ are the averages of $x$ and $y$, respectively, $\sigma _{x}^{2}$ and $\sigma _{y}^{2}$ are the variances of $x$ and $y$, respectively, $\sigma _{xy}$ is the covariance of $x$ and $y$, and $c_{1}$ and $c_{2}$ are two hyperparameters used to stabilize the division operation.
\subsection{Results}

\subsubsection{Module structure}
To determine the most effective alternative network for our purposes, a series of experiments were conducted on the backbone network. The alternative networks were U-Net \cite{ronneberger2015u}, R2U-Net \cite{alom2018recurrent}, AttU-Net \cite{oktay2018attention}, R2AU-Net \cite{zuo2021r2au}, CPCE \cite{shan20183} and ResNet \cite{he2016deep}. Each of them was integrated as a base module into the overall process, sharing the same structure and parameters.

At this stage of research, the entire network consisted of 5 modules. The employed noise model had a mean value of 0 and a variance of 15. These specifications set the foundation for evaluating the performance of the alternative networks.
The results obtained from the testing set for each backbone network are presented in Table \ref{t1}. Among the alternative networks, U-Net demonstrated the best performance in terms of both the PSNR and SSIM. R2U-Net, AttU-Net, and R2AU-Net did not achieve better results despite the addition of the residual module, recurrent module, and self-attention module to U-Net, respectively. CPCE and ResNet achieved similar PSNR results. CPCE performed well in terms of the SSIM but was not the best model. This outcome established U-Net as the most effective backbone network for the given task.

\begin{figure*}
	\centering
	\includegraphics[width=0.7\textwidth]{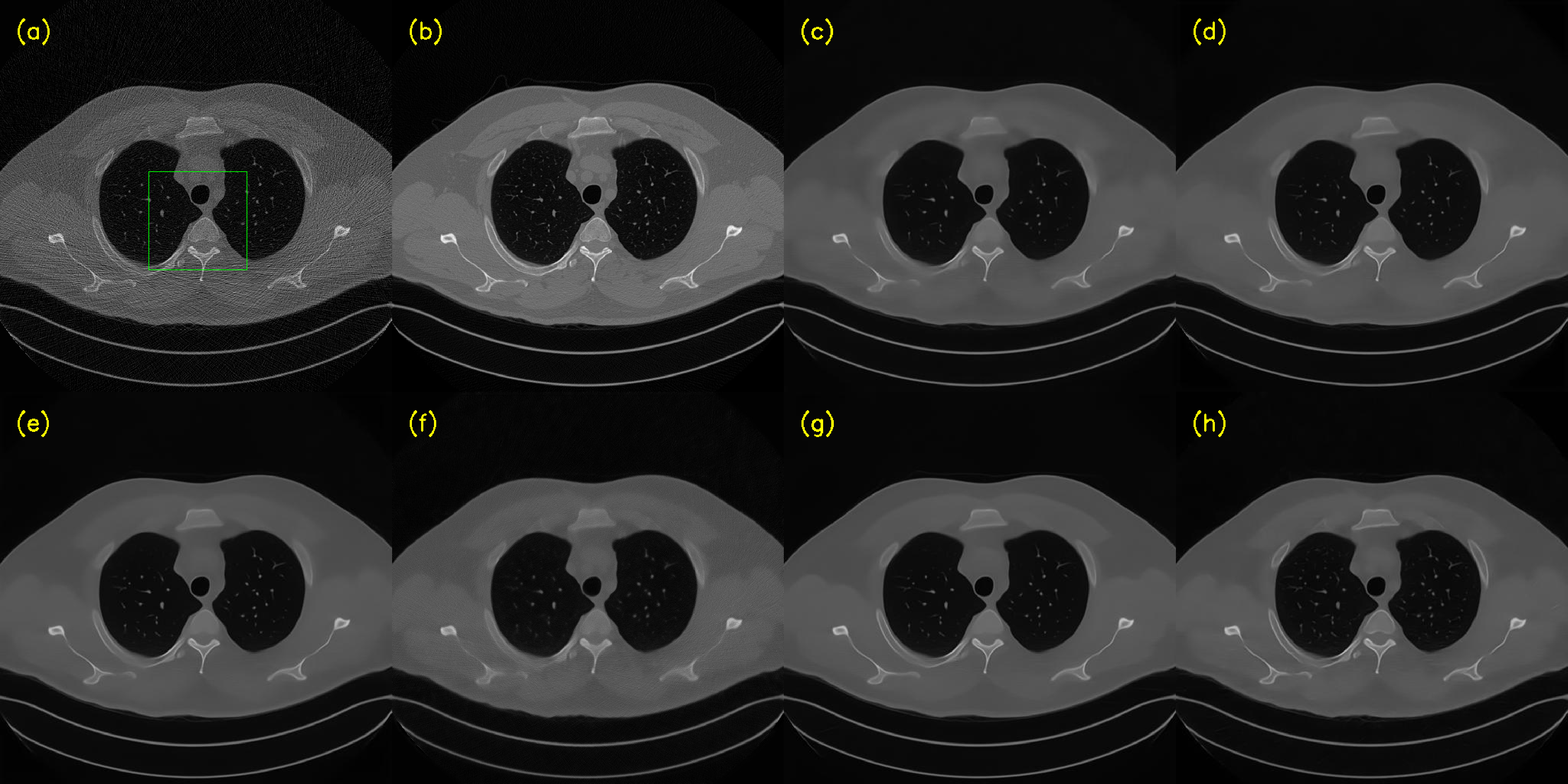}
\caption{Visual comparison among the denoising results obtained by different backbone networks. Among them, (a)-(h) represent the LDCT image, the NDCT image, R2U-Net, AttU-Net, R2AttU-Net, CPCE, ResNet, and U-Net (ours).}
	\label{figure3}
\end{figure*}

\begin{figure*}
	\centering
	\includegraphics[width=0.7\textwidth]{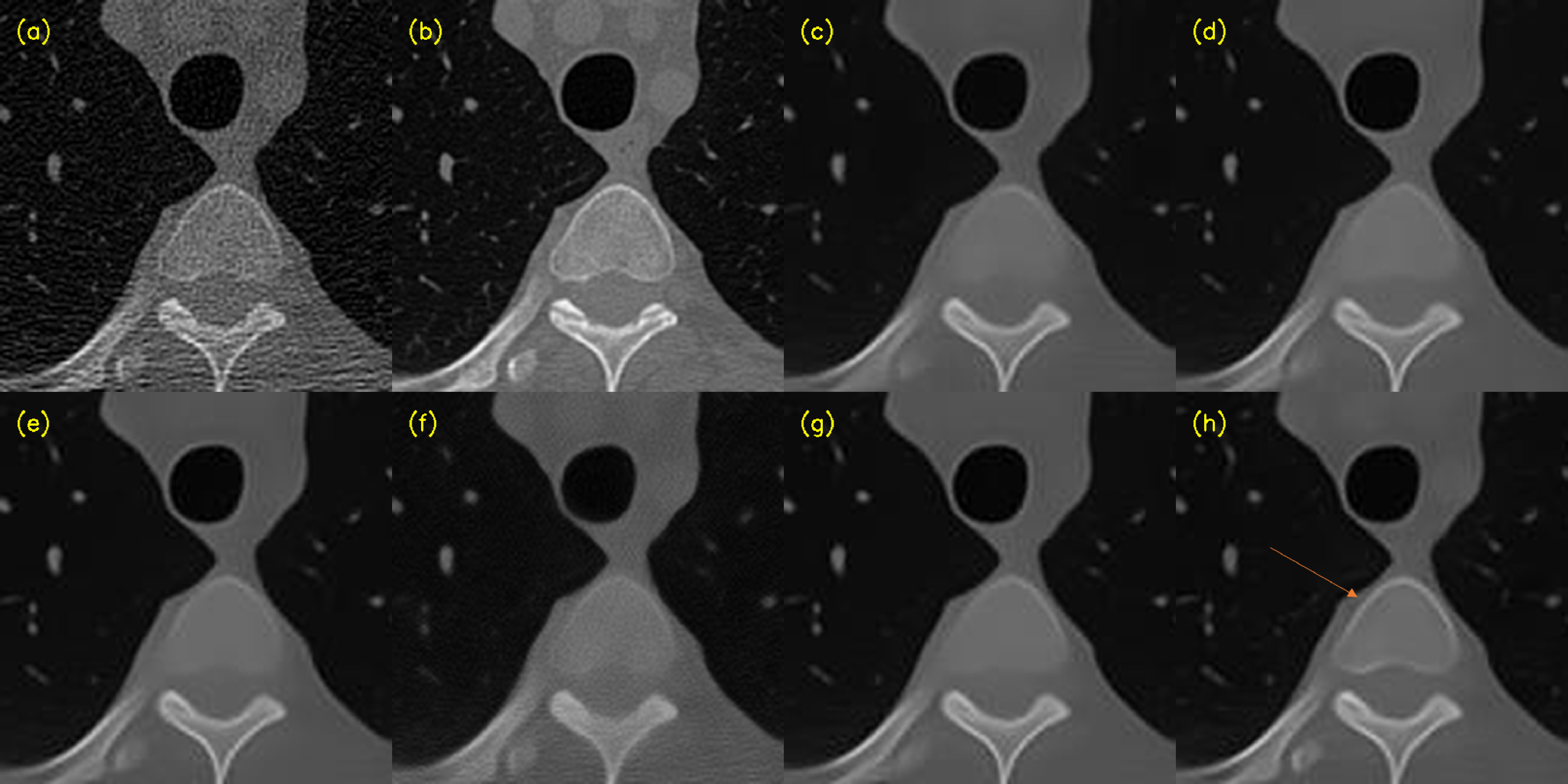}
\caption{The zoomed ROI regions represented by the green rectangles of Figure \ref{figure3}. The orange arrows show the parts of the mottled areas restored by U-Net.}
	\label{figure3_2}
\end{figure*}

To provide a visual comparison among the denoising results achieved by different backbone networks, Figure \ref{figure3} was generated. Additionally, Figure \ref{figure3_2} focuses on an enlarged portion of the region of interest (ROI). It is evident that employing U-Net as the backbone network not only led to effective denoising but also achieved an enhanced tissue detail preservation effect. U-Net smoothed out most of the tissue details while successfully recovering some mottled areas to a certain extent, as indicated in Figure \ref{figure3_2} (h). Overall, U-Net exhibited superior denoising and detail preservation capabilities to those of the other networks.

Consequently, based on the obtained results, U-Net was chosen as a module for integration into our proposed modular neural network for subsequent experiments.
\begin{table}[width=.9\linewidth,cols=4,pos=h]\label{tab:exist}
\begin{center}
        
        \caption{Results obtained by different backbone networks on the testing set.}
        \label{t1}
	\begin{tabular}{ccc} 
	\hline
	& PSNR & SSIM \\ 
	\hline
	R2U-Net & 24.547 & 0.6509\\ 
	AttU-Net & 24.977 & 0.6800\\
        R2AU-Net & 24.559 & 0.6773\\
        CPCE & 24.765 & 0.6898\\
        ResNet & 24.765 & 0.6697 \\
        \textbf{U-Net (Ours)} & \textbf{27.294} & \textbf{0.6978} \\
	\hline
\end{tabular}
\end{center}
\end{table}

\subsubsection{Number of modules}

\begin{figure*}
	\centering
	\includegraphics[width=1\textwidth]{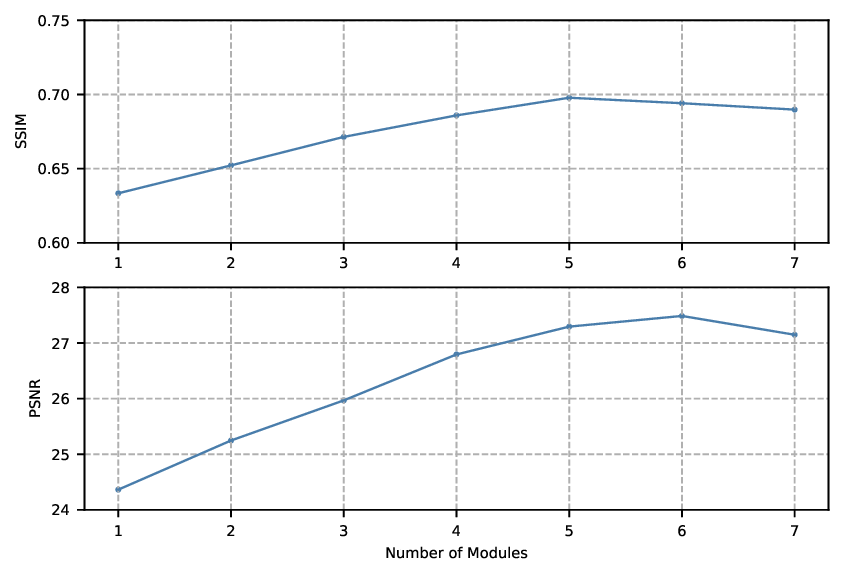}
	\caption{Results obtained on the testing set with different numbers of modules.}
	\label{exp2}
\end{figure*}

The determination of the optimal number of modules ($T$) was carried out through a series of experiments. Figure \ref{exp2} presents the results obtained when varying the number of modules from 1 to 7. The performance of the model demonstrated an increasing trend as the number of modules began to increase, up to a threshold of 5. Beyond this point, the curve declined.

These findings indicate that the effect increased in an approximately linear manner when the number of modules was less than 5. Increasing the number of modules beyond this range not only failed to yield significant performance gains but also introduced distortions in the denoising results. Consequently, we selected 5 as the number of modules to ensure both effective denoising and high computational efficiency.

\subsubsection{Noise level}
To investigate the impact of different noise levels on the experimental results, we conducted experiments by introducing noise with various intensities. The objective was to determine the optimal effect of the chosen noise. The mixed noise model employed in this study combined Poisson and Gaussian noise. Since the Poisson noise is only related to the pixel value, we next discuss zero-mean Gaussian noise in the context of mixed noise. We specifically examined the effects of adding Gaussian noises with different variances to the experimental setup to gain insights into their influences on the outcomes.

\begin{figure*}
	\centering
	\includegraphics[width=1\textwidth]{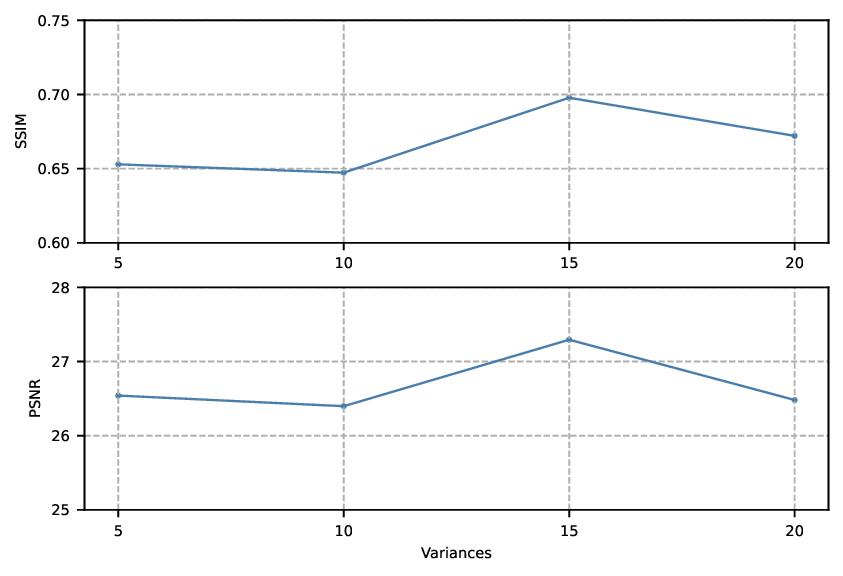}
	\caption{Results obtained by noise models with different variances on the testing set.}
	\label{exp3}
\end{figure*}

Figure \ref{exp3} presents the results and the corresponding trends observed when adding noise models with different variance values to the training set. It is worth noting that the results did not exhibit a linear increase with increasing noise variance. Instead, the model performed relatively well when the variance was set to 15.
These findings indicate that the impact of different noise levels on the experimental results was not solely dependent on the intensity or variance of the added noise. Instead, there seemed to be an optimal level at which the model performed most effectively. In this case, the variance value of 15 appeared to yield the best results. As follows, we chose to use a mixture of zero-mean Poisson noise and zero-mean Gaussian noise with a variance of 15 as the noise model.

\subsubsection{Ablation experiments}

To validate the accuracy of the noise model and the feasibility of the noisy-as-clean strategy, a series of ablation experiments were conducted. The purpose was to assess the individual contributions and effectiveness levels of these components. To accomplish this task, specific modifications were made to the experimental setup.

First, to assess the validity of the mixed noise model, we replaced it with independent noise. Zero-mean Poisson noise (or Gaussian noise) corrupted the data multiple times, contrasting with mixed noise. This allowed for a comparison between the performance achieved by the model with the Poisson noise or Gaussian noise and that achieved with the mixture of Poisson and Gaussian noise.
Next, the noisy-as-clean strategy was examined by eliminating its implementation. Instead, the training process was conducted by repeatedly adding noise to the NDCT dataset. The target was to determine the importance and effectiveness of the noisy-as-clean strategy for achieving accurate results.

The outcomes of these ablation experiments are presented in Table \ref{table4}, which displays the corresponding testing results obtained from each configuration. It was observed that the noise model exerted substantial influence on the final outcome, indicating its necessity and effectiveness in the model.
Furthermore, when comparing the model's performance with LDCT and NDCT images as training sets, no significant differences were observed. This observation provided evidence that supported the validity of the noisy-as-clean strategy. This approach obviated the need for using NDCT images, which are more challenging datasets to obtain.
\begin{table}[htbp]
	\begin{center}
		
		\caption{Results of ablation experiments involving the noise model and the noisy-as-clean strategy ('w' means 'with', and 'w/o' means 'without').}
            \label{table4}
		\begin{tabular}{ccc} 
			\hline
			& PSNR & SSIM \\ 
			\hline
			Ours w/o noise model (w Poisson noise) & 23.153 & 0.5578\\ 
   Ours w/o noise model (w Gaussian noise) & 22.418 & 0.5233\\ 
 Ours w/o noisy-as-clean
 strategy & \textbf{27.384} & 0.6898\\
\textbf{Ours}&27.294 &\textbf{0.6978}\\
			\hline
		\end{tabular}
	\end{center}
\end{table}


\subsubsection{Method comparison}
To comprehensively evaluate the overall effectiveness of our network, we compared it with seven available state-of-the-art denoising methods. These methods encompassed a range of approaches, including traditional denoising, self-supervised, and unsupervised methods. Specifically, the evaluated methods were BM3D \cite{dabov2006image}, N2V \cite{krull2019noise2void}, NBR2NBR \cite{huang2021neighbor2neighbor}, B2U \cite{wang2022blind2unblind}, a CycleGAN \cite{zhu2017unpaired}, an ADN \cite{liao2019adn}, and a Dual-GAN \cite{yi2017dualgan}.

BM3D represents a classic image denoising method. The CycleGAN, ADN, and Dual-GAN were trained using both LDCT and NDCT images. In contrast, N2V, NBR2NBR, B2U, and our model were trained exclusively on LDCT images, using the same number of slices as the aforementioned methods but without the utilization of NDCT data. It is worth noting that our method did not use the traditional NDCT data as the training set but rather LDCT images. This meant that a good noise reduction effect could be achieved without the need for high-risk NDCT images.

Table \ref{table5} provides an overview of the denoising results obtained by each method on the testing set. Our method outperformed the traditional denoising method (BM3D) by a significant margin. Among the self-supervised methods that do not rely on NDCT images, our method achieved the highest score, indicating its superior denoising capabilities. Although our method achieved a slightly lower PSNR than that of the CycleGAN, it exhibited a higher SSIM. This implies that our method may possess a slightly weaker denoising capability but better preserves the structural details of images than the CycleGAN. Furthermore, our method surpassed the other unsupervised methods, such as the ADN and Dual-GAN. Surprisingly, the Dual-GAN did not demonstrate strong performance in this comparison.

\begin{figure*}
	\centering
	\includegraphics[width=0.9\textwidth]{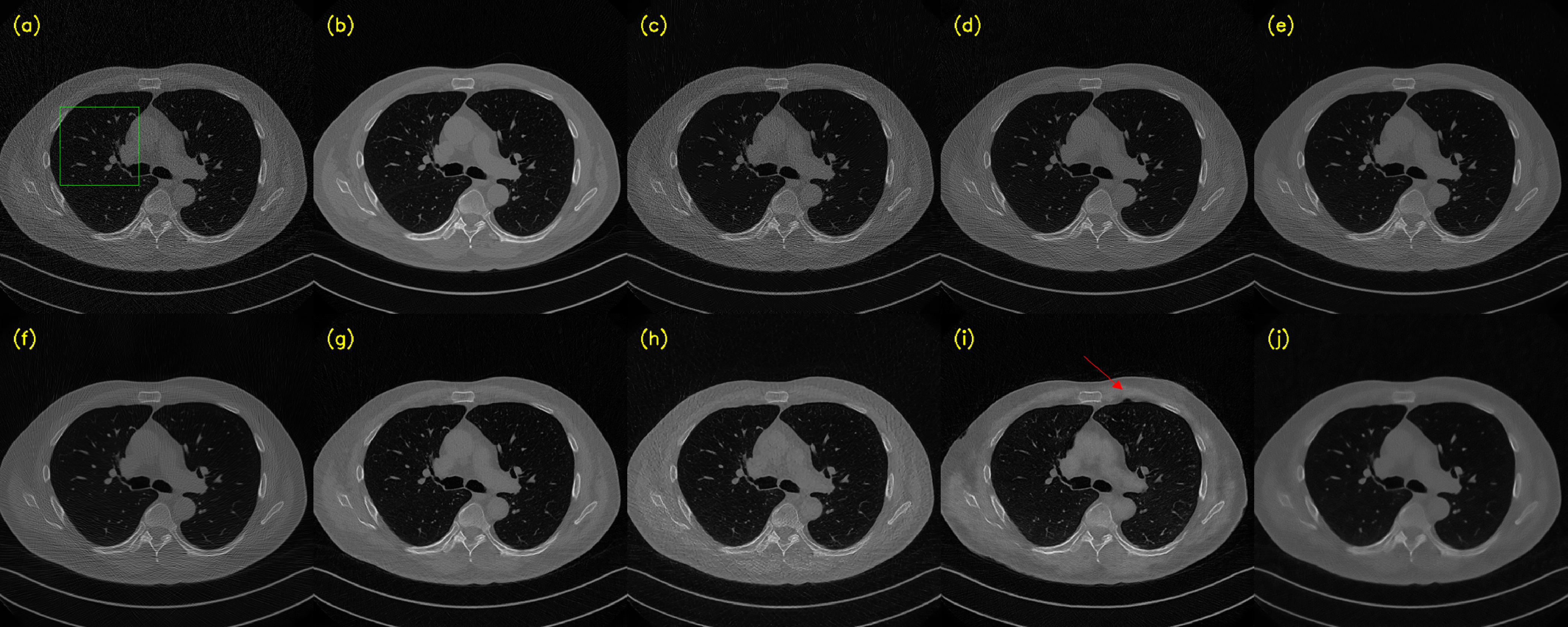}
	\caption{Comparison of chest sections using different methods. Among them, (a)-(j) LDCT, NDCT, BM3D, N2V, NBR2NBR, B2U, CycleGan, ADN, Dual-GAN and ours.}
	\label{figure7}
\end{figure*}

\begin{figure*}
	\centering
	\includegraphics[width=0.9\textwidth]{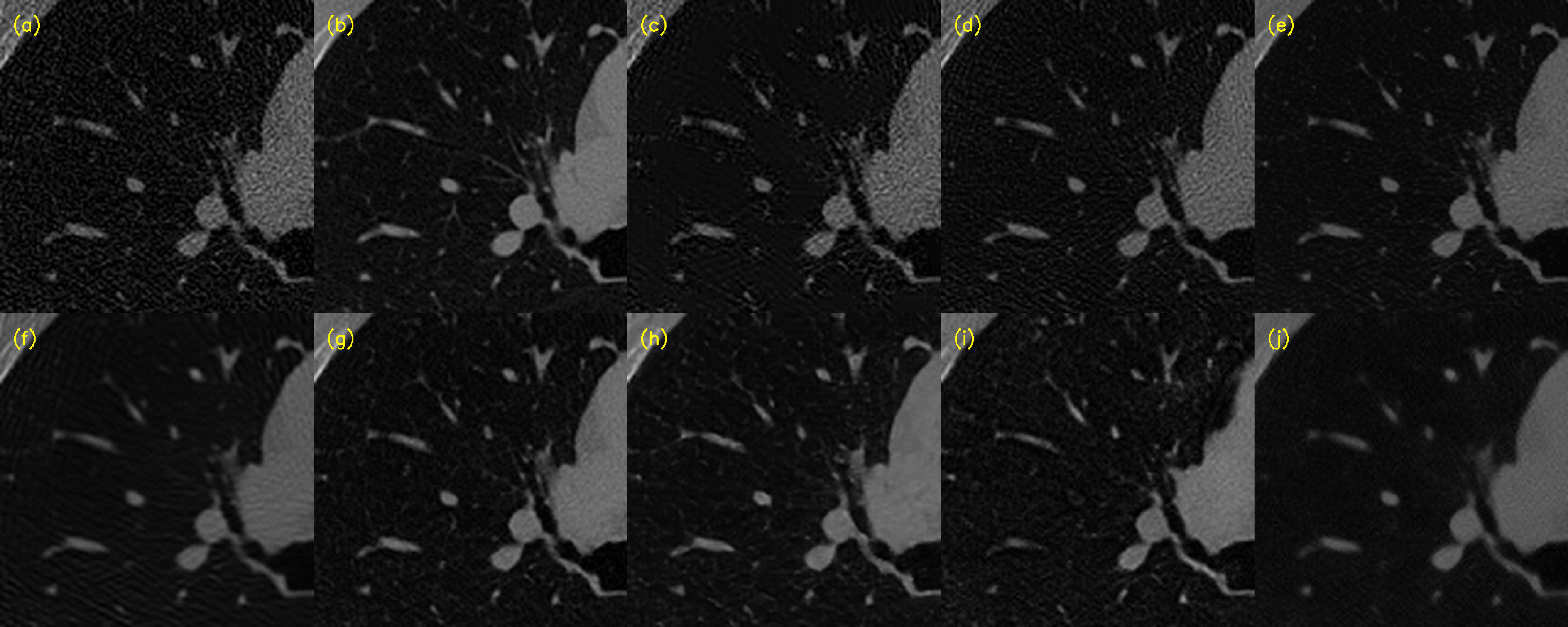}
	\caption{The zoomed ROI regions represented by the green rectangles of Figure \ref{figure7}. The red arrow in (i) highlights tissue distortion exhibited by the Dual-GAN.}
	\label{figure7_2}
\end{figure*}

Figure \ref{figure7} provides a visual comparison of the chest slice results obtained using different mainstream denoising methods. To provide a more detailed analysis, Figure \ref{figure7_2} presents an enlarged view of the ROI. Our method produced visually appealing results, effectively removing a significant portion of the noise and structural artifacts while retaining a considerable degree of structural detail. Furthermore, our method produced a notably smoother output than those of the other methods. However, the results of the other self-supervised methods retained some noticeable noise. The results of the unsupervised methods introduced some global, structural noise, as seen, for example, in the tissue distortions produced by the Dual-GAN.

In summary, the comparative evaluation demonstrated that our network outperformed traditional methods such as BM3D, exhibited competitive performance when compared to the CycleGAN, surpassed other self-supervised methods and unsupervised methods, and achieved the best results among the self-supervised methods that did not utilize NDCT data. The visual analysis also reinforced the effectiveness of our method, showcasing its ability to effectively reduce noise, preserve structural details, and yield visually appealing results.
\vspace*{2ex}
\begin{table}[htbp]
	\begin{center}
		
		\caption{Results of different denoising methods.}
            \label{table5}
		\vspace*{2ex}
		\begin{tabular}{ccc} 
			\hline
			& PSNR & SSIM \\ 
			\hline
			LDCT & 21.698 & 0.4176\\ 
			BM3D & 22.027& 0.5133 \\
			N2V&25.486&0.6298\\
			NBR2NBR&24.112&0.6363\\
			B2U&25.818&0.5828\\
			CycleGAN&\textbf{27.965}&0.6926\\
			ADN&27.204&0.5767\\
			Dual-GAN&24.544&0.5868\\
			\textbf{Ours}&27.294&\textbf{0.6978}\\
			\hline
		\end{tabular}
	\end{center}
\end{table}

\section{Conclusions}

This paper proposed a combination including the self-supervised N2N model and the noisy-as-clean strategy for LDCT denoising with a modular network structure to address the difficulty of obtaining high-quality paired or unpaired NDCT data. In the methodology section, we verified the theoretical validity of the important parts of the network structure. In the experimental section, we determined the backbone network, the noise level, and the number of modules. Then, the effectiveness of our method and the necessity of each part of its components were also determined through ablation experiments. Our approach outperformed all self-supervised methods and most unsupervised methods, demonstrating its superiority.

With the help of the mixed noise model, we obtained good results. However, the variance of the noise model still needed to be manually and experimentally determined. In future work, we will explore how to adaptively determine the strength of the noise model to fit more data.

\section*{Acknowledgments}
	
This paper was supported by the Natural Science Foundation of Liaoning Province (2022-MS-114).

\bibliographystyle{vancouver} 
\bibliography{cas-sc-template}
\end{sloppypar}
\end{document}